\documentclass{appolb}
\usepackage{graphicx,cite}
\usepackage{amssymb}
\usepackage{upgreek}
\usepackage{color}
\def\ifm#1{\relax\ifmmode#1\else$#1$\fi}  \def\to{\ifm{\rightarrow}}  \def\epm{\ifm{e^+e^-}}
\def\gam{\ifm{\gamma}}     \def\x{\ifm{\times}}  \def\ab{\ifm{\sim}}    
\def\up#1;{$^{#1}$}  \def\dn#1;{$_{#1}$}    \def\DAF{DA\char8NE}  \def\f{\ifm{\phi}}
\def\Li{\ifm{L_i}}  \def\pic{\ifm{\pi^+\pi^-}}  \def\po{\ifm{\pi^0}}
\def\pt#1;#2;{\ifm{#1\x10^{#2}}}  
\def\mt{\ifm{M_{\rm trk}}}

\begin{document}
% \eqsec  % uncomment this line to get equations numbered by (sec.num)
\title{
Search for the U-boson in the process $e^ +e^- \to \mu^+ \mu^- \gamma, \,\, U \to \mu^+\mu^-$ with the KLOE detector
%
%\thanks{Presented at ...}%
% you can use '\\' to break lines
}
\author{Francesca Curciarello
\address{Dipartimento di Fisica e di Scienze della Terra dell'Universit\`a di Messina and INFN Sezione di Catania, Italy}
\\
\address{on behalf of the KLOE-2 collaboration \thanks{
The KLOE-2 Collaboration:
D.~Babusci,
%\author[Roma2,INFNRoma2]{D.~Badoni},
I.~Balwierz-Pytko,
G.~Bencivenni,
%\author[Roma1,INFNRoma1]{C.~Bini},
C.~Bloise,
%\author[INFNRoma1]{V.~Bocci},
F.~Bossi,
P.~Branchini,
A.~Budano,
%\author[Moscow]{S.~A.~Bulychjev},
L.~Caldeira~Balkest\aa hl,
%\author[Frascati]{P.~Campana},
G.~Capon,
F.~Ceradini,
P.~Ciambrone,
F.~Curciarello,
E.~Czerwi\'nski,
E.~Dan\`e,
V.~De~Leo,
E.~De~Lucia,
G.~De~Robertis,
A.~De~Santis,
%\author[Roma1,INFNRoma1]{G.~De~Zorzi},
P.~De~Simone,
A.~Di~Cicco,
A.~Di~Domenico,
C.~Di~Donato,
R.~Di~Salvo,
%\author[Roma3,INFNRoma3]{B.~Di~Micco},
D.~Domenici,
%\author[Frascati]{A.~D'Uffizi},
O.~Erriquez,
G.~Fanizzi,
A.~Fantini,
G.~Felici,
S.~Fiore,
P.~Franzini,
A.~Gajos,
P.~Gauzzi,
G.~Giardina,
S.~Giovannella,
%\author[Roma2,INFNRoma2]{F.~Gonnella},
E.~Graziani,
F.~Happacher,
L.~Heijkenskj\"old,
B.~H\"oistad,
%\author[Frascati]{L.~Iafolla},
%\author[Energetica,Frascati]{E.~Iarocci},
M.~Jacewicz,
T.~Johansson,
K.~Kacprzak,
D.~Kami\'nska,
%\author[Warsaw]{A.~Kowalewska},
%\author[Moscow]{V.~Kulikov},
A.~Kupsc,
J.~Lee-Franzini,
%\author[Frascati]{B.~Leverington},
F.~Loddo,
S.~Loffredo,
G.~Mandaglio,
M.~Martemianov,
M.~Martini,
M.~Mascolo,
%\author[Moscow]{M.~Matsyuk},
R.~Messi,
S.~Miscetti,
G.~Morello,
D.~Moricciani,
P.~Moskal,
F.~Nguyen,
%\author[Frascati]{L.~Quintieri},
A.~Palladino,
A.~Passeri,
V.~Patera,
I.~Prado~Longhi,
A.~Ranieri,
%\author[Mainz]{C.~F.~Redmer},
P.~Santangelo,
I.~Sarra,
M.~Schioppa,
B.~Sciascia,
%\author[Energetica,Frascati]{A.~Sciubba},
M.~Silarski,
%\author[Calabria,INFNCalabria]{S.~Stucci},
C.~Taccini,
L.~Tortora,
G.~Venanzoni,
%\author[Frascati,CERN]{R.~Versaci},
W.~Wi\'slicki,
M.~Wolke,
J.~Zdebik}}
}

\maketitle
\begin{abstract}
We present a search for a new light vector boson, carrier of a ``dark force'' between WIMPs, with the KLOE detector at \DAF. We analysed $e^+ e^- \to \mu^+ \mu^- \gam$ ISR events corresponding to an integrated luminosity of $239$~pb$^{-1}$ to find evidence for the  $e^+ e^- \to U\gam ,\,\, U\to\mu^+\mu^-$ process.
We found no $U$ vector boson signal and set a 90\%  CL upper limit on the ratio of the U boson and photon coupling constants between 1.6$\x$10$^{-5}$ to 8.6$\x$10$^{-7}$ in the mass region $520<M_{\rm U}<980$ MeV. A projection of the KLOE sensitivity for the $\mu\mu\gamma$ and $\pi\pi\gamma$ channels at full statistics and extended muon acceptance is also presented.
\end{abstract}
\PACS{PACS numbers come here}
  
\section{Introduction}

Many extensions of the Standard Model (SM)~\cite{Holdom,U_th1, Fayet,U_th2,U_th6} assume that dark matter (DM) is made up of new particles charged under some new interaction mediated by a new gauge vector boson called $U$ (also referred to as dark photon or A'). The $U$ boson can kinetically mix with the ordinary photon through 
high-order diagrams, providing therefore a small coupling with SM particles~\cite{Holdom,U_th1,U_th2,Fayet,U_th6}. 
The coupling strength can be expressed by a single factor, $\varepsilon$, equal to the ratio of dark
and Standard Model electromagnetic couplings~\cite{Holdom}. 
A $U$ boson with mass of $\mathcal{O}$(1GeV) and $\varepsilon$ in the range 10\up-2;--10\up-7; could
explain all puzzling effects observed in recent astrophysics experiments~\cite{Pamela,AMS,Integral,Atic,Hess,Fermi,Dama/Libra}. By using a data sample corresponding to an integrated luminosity of 239~pb\up-1;, KLOE investigated the  the radiative $U$\gam \, production with $U\to\mu^+\mu^-$. New searches are foreseen to exploit the full KLOE statistics for the $\mu^+\mu^-\gamma$ channel and also to search for the $U\to\pi^+\pi^-$ decay.
%We employed KLOE data collected in 2002 at \DAF \, \epm\  collider, running at the \f-meson peak, with an integrated luminosity of 239.3 pb\up-1;. We decided to investigate the radiative $U$\gam \, production limiting our search to the muon decay channel ($U\to\mu^+\mu^-$) which has with available statistics a reach in sensitivity of 10\up-3;--10\up-2;, for $U$-boson masses, $M_{\rm U}$,  up to a few GeV~\cite{U_th1,Fayet,U_th2,U_th6}. The $U$-boson peak would appear in the dimuon mass spectrum. 

\section{The KLOE detector}
The KLOE detector operates at DA$\Phi$NE, the Frascati $\phi$-factory. 
It consists of a large cylindrical drift chamber (DC)~\cite{KLOE_DC}, surrounded by a lead scintillating-fiber electromagnetic calorimeter (EMC)~\cite{KLOE_EMC}.  
A superconducting coil around  the EMC provides a 0.52 T magnetic field along the beam axis.
EMC energy and time resolutions are $ \sigma_E /E\,=\,0.057/\sqrt{E(\rm{GeV})} $ and $ \sigma_t =57\ \rm{ps}/\sqrt{E(\rm{GeV})}\oplus 100\ \rm{ps}$, respectively. 
The drift chamber has only stereo sense wires, it is $4$ m in diameter, $ 3.3$ m long and operates with a low-$Z$ gas mixture (helium with 10\% isobutane). Spatial resolutions are  $\sigma_{xy}\ab150\ \rm\upmu m$ and  $\sigma_z\ab2$ mm. The momentum resolution for large angle tracks is $\sigma(p_\perp) / p_\perp\ab 0.4\% $.

\section{$\mu^+ \mu^- \gam$ data analysis}\label{Data analysis}

The $\mu^+ \mu^- \gam$ event selection requires two tracks of opposite charge with $50^\circ\!<\!\theta\!<\!130^\circ$ and an undetected photon whose momentum, computed according to the $\mu \mu \gamma$ kinematics, points at small polar angle ($\theta<15^\circ,>~165^\circ$)\cite{KLOE_U,KLOE_pi_FF}. These requirements limit the range of $M_{\mu \mu}$ to be larger than 500~MeV and greatly reduce the contamination from the resonant  and Final State Radiation (FSR) processes: $\epm\to\phi\to\pi^+\pi^-\pi^0$, $\epm\to \pi^+\pi^-\gamma_{\,\rm{FSR}}$ and $\epm\to \mu^+\mu^-\gamma_{\,\rm{FSR}}$.
The above selection criteria are also satisfied by $\epm\to\epm\gam$ radiative Bhabha events. To obtain additional separation between electrons and pions or muons, a particle identification estimator (\Li), based on a pseudo-likelihood function using time-of-flight and calorimeter information is used~\cite{KLOE_U,KLOE_pi_FF}. Events with both tracks satisfying $\Li\!<\!0$ are rejected as $\epm\gam$ with a $\pi\pi\gamma/\mu\mu\gamma$ loss less than 0.05\%. 
\begin{figure}[htp!]
\begin{center}
\includegraphics[width=6cm]{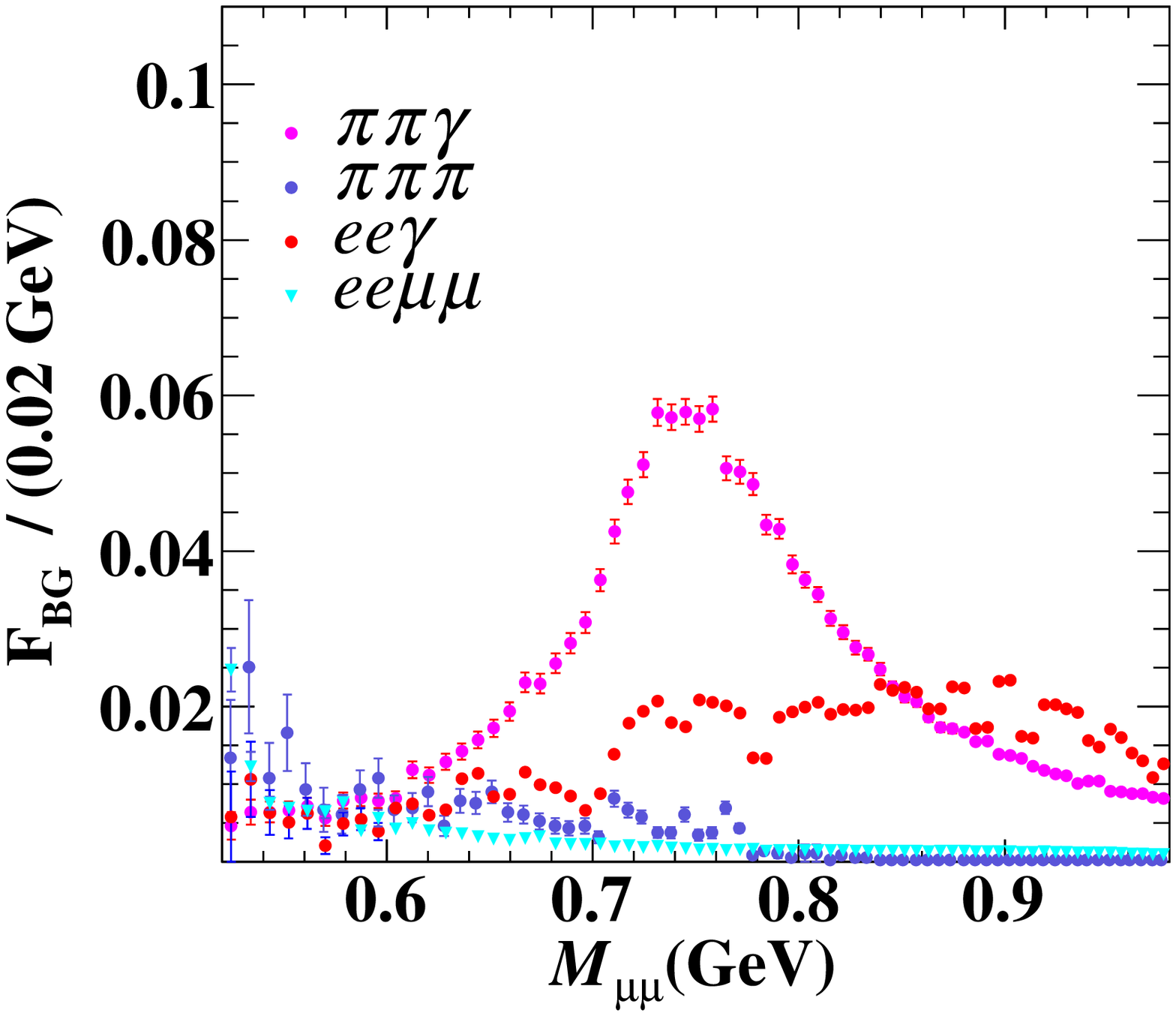}
\includegraphics[width=5.8cm]{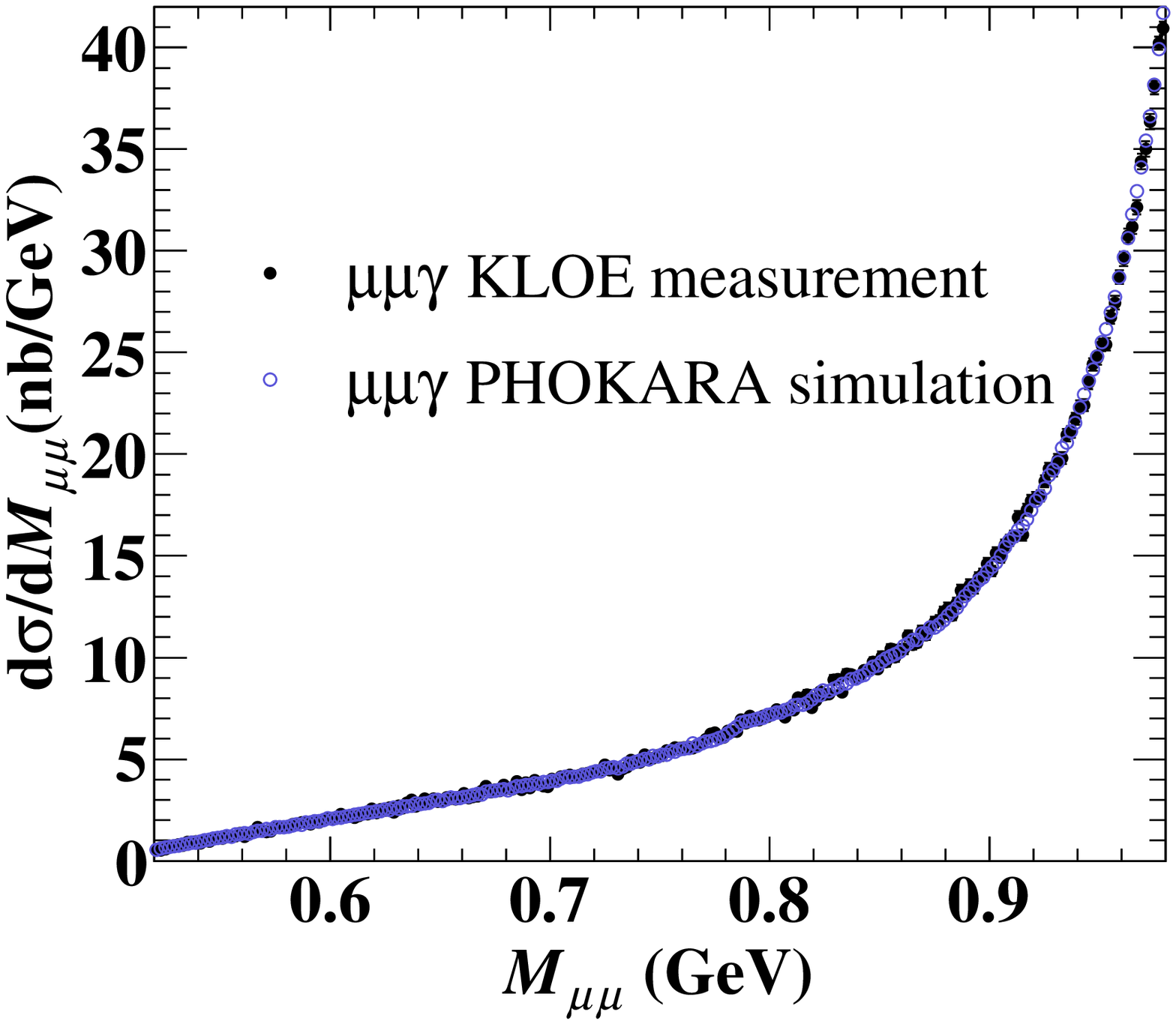}
\caption{Left: Fractional backgrounds to the $\mu\mu\gam$ signal; see insert for symbols. Right: Comparison of data (full circles) and simulation (open circles) for $\mu^+ \mu^- \gamma$ cross section.}
\label{Bckg}
\end{center}
\end{figure}
Pions and muons are identified by means of the variable $M_{\rm trk}$ defined as the mass of oppositely-charged particles in the $\epm\to x^+ x^- \gam$ process in which we assume the presence of an unobserved photon and that the tracks belong to particles of the same mass with momentum equal to the observed value.
The $M_{\rm trk}$ values between  80--115 identify muons while $M_{\rm trk}$ values $>$130 MeV identify pions. A cut on the quality of the fitted tracks, parametrized by the error on \mt, $\sigma_{M_{\mathrm{trk}}}$, has been implemented to further improve the $\pi/\mu$  separation~\cite{KLOE_U}.
%  \begin{figure}[htp!]
%\begin{center}
%\includegraphics[width=5.8cm]{fig4a.eps}
%\includegraphics[width=5.8cm]{fig4b.eps}
%\caption{Left: Comparison of data (full circles) and simulation (open circles) for $\mu^+ \mu^- \gamma$ cross section; Right: Ratio of the two spectra fitted with a constant function.}
%\label{mmg_abs}
%\end{center}
%\end{figure}
At the end of the analysis chain, residual backgrounds consisting of \epm\to\epm\gam, \epm\to\pic\gam\  and  \epm\to\f\to\pic\po\ are still present. The last two contributions have been evaluated from Monte Carlo (MC) simulation while the \epm\to\epm\gam\, events have been estimated directly from data ~\cite{KLOE_U,KLOE_pi_FF}.
Additional background from \epm\to\epm$\mu^+\mu^-$  is at the percent level below 0.54 GeV and decreases with $M_{\mu\mu}$. 
Fig.~\ref{Bckg} left shows the fractions of the background processes, $F_{\rm BG}$, as a function of $M_{\mu\mu}$ while Fig.~\ref{Bckg} right shows the measured $\mu\mu\gamma$ cross section compared with the NLO QED calculations, using the MC code PHOKHARA~\cite{PHOKHARA}.  The  agreement between measurement and the PHOKHARA simulation is excellent, proving analysis consistency. No structures are visible in the $M_{\mu\mu}$ spectrum.

\section{\mathversion{bold} Limit on $U$-boson coupling and future prospects }
\label{UL}

To exclude small  $U$-boson signals  we extracted the limit on the number of $U$-boson candidates at 90\% of confidence level (CL) through the CLS technique~\cite{CLS_Technique}. We compared the expected and observed $\mu^+\mu^- \gam$ yield, and a MC generation of the $U$-boson signal which takes into account the resolution in $M_{\mu\mu}$ (from 1.5~MeV to 1.8~MeV as $M_{\mu\mu}$ increases).
\begin{figure}[htp!]
\begin{center}
 \includegraphics[width=8.5cm]{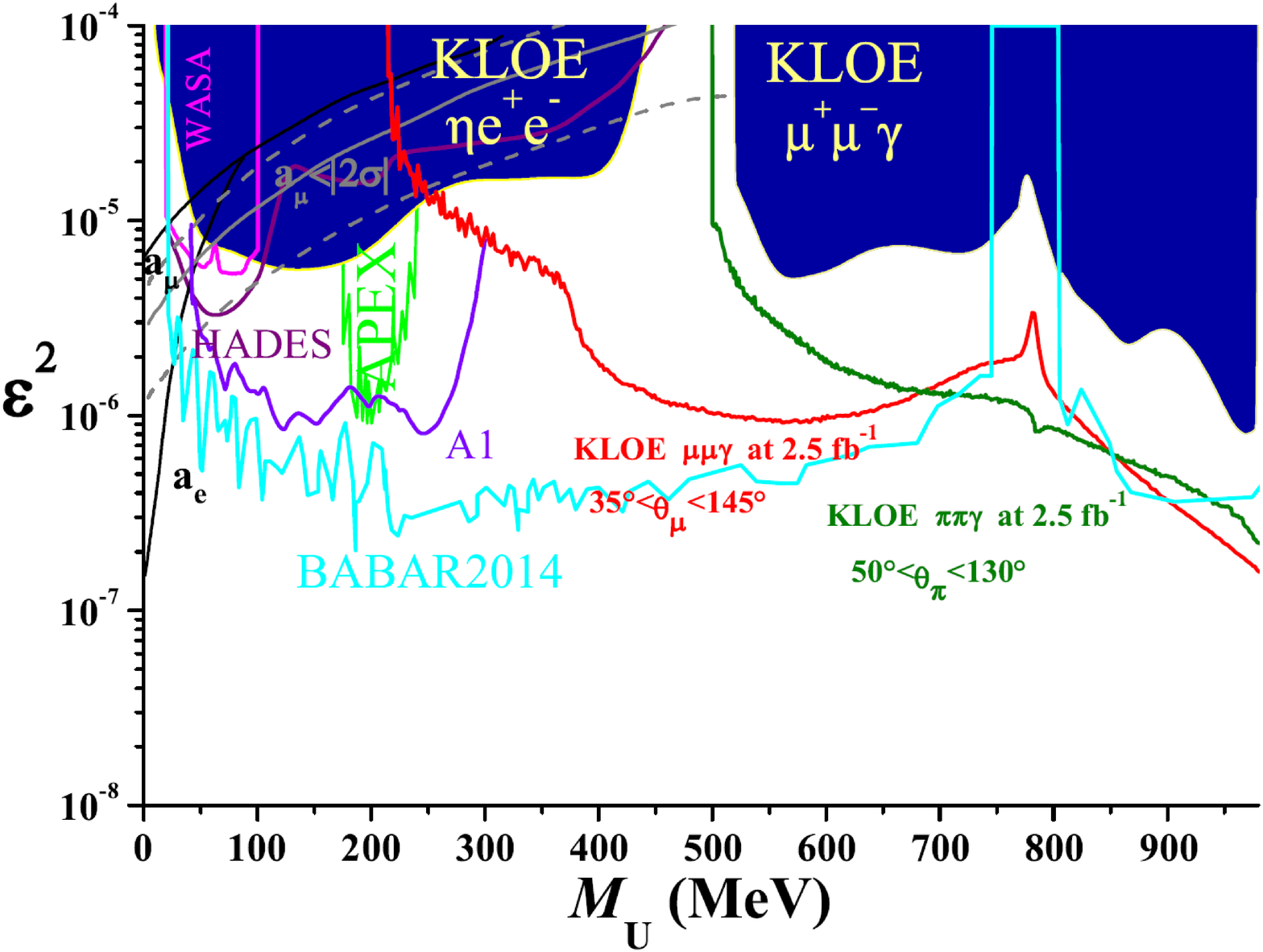}
\caption{ 90\% CL exclusion plot  for $\varepsilon^2$ as a function of the $U$-boson mass (blue)~\cite{KLOE_U}. Limits from  A1~\cite{Mami1} (violet), Apex~\cite{Apex} (green), KLOE $\phi\to\eta\epm$~\cite{KLOE_UL1,KLOE_UL2} (blue), WASA~\cite{WASA} (magenta), HADES~\cite{HADES}  (purple) and BaBar (cyan)~\cite{BaBar} are shown. The black and grey lines are the limits from the muon and electron anomaly~\cite{a_mu}, respectively. KLOE sensitivity for $\mu\mu\gamma$ and $\pi\pi\gamma$ at 2.5~fb$^{-1}$ and $35^{\circ}(50^{\circ})<\theta_{\mu(\pi)}<145^{\circ}(130^{\circ})$ are also shown (red and green lines, respectively).}
%\label{Exclusion_plot}
\label{Raw_TLimit_Output}
\end{center}
\end{figure}
The limit on the kinetic mixing parameter has been extracted by means of:%~\cite{memo_U}:
\begin{equation}
 \varepsilon^2= \frac{N_{\rm{CLS}
  }/ (\epsilon_{\rm{eff}} \x L)}{H \x I},
\label{eq.3}
\end{equation}
where  $\epsilon_{\rm{eff}}$ represents
the overall efficiency (1-15\% as $M_{\mu\mu}$ increases~\cite{KLOE_U}), $L$ is the integrated luminosity, $H$ is the radiator function obtained from QED including NLO corrections, and $I$ is the effective $U$ cross section~\cite{Fayet}. 
The resulting exclusion plot on the kinetic mixing parameter
$\varepsilon^2$, in the 520--980~MeV mass range,  is shown in
Fig.~\ref{Raw_TLimit_Output}.
In the same plot,  other limits in the mass range below
$1$ GeV are also shown~\cite{Mami1,Apex,WASA,KLOE_UL1,KLOE_UL2,BaBar,HADES,a_mu}.
Our 90\% CL limit is between 1.6$\x$10$^{-5}$ and 8.6$\x$10$^{-7}$ in the
520--980~MeV mass range~\cite{KLOE_U}.

%\section{Projection of KLOE sensitivity for the $\mu\mu\gamma$ channel}
%\label{prospects}
An upgrade of the presented analysis is foreseen by employing the full KLOE data statistics corresponding to an integrated luminosity of 2.5~fb$^{-1}$  and by extending the muon polar angle acceptance from 50$^{\circ}$ to 35$^{\circ}$ and from 130$^{\circ}$ to 145$^{\circ}$ . The KLOE reach in sensitivity by considering a 2~MeV invariant mass resolution is presented in Fig.~\ref{Raw_TLimit_Output}  (red line) at $N_{\rm U}/\sqrt{N_{\rm QED}}=2$. 
%The great difference with the actual limit is due to systematic errors  and background contributions added to the data spectrum for the limit extraction that decrease the KLOE reach at 240pb$^{-1}$. 
%%The experimental limit at full statistics should be thus less stringent than expected one of a factor of about four. 
The sensitivity loss due to the $\rho$ meson around 770~MeV  is due to the branching fraction of the $U \to \mu^+ \mu^-$ channel which is suppressed by the dominant hadronic decay mode $U \to \pi^+ \pi^-$. To overcome this problem KLOE-2 plans to carry on also a new analysis by exploiting the $\pi^+ \pi^- \gamma$ channel. The detailed KLOE-2 physics program is presented in Ref.~\cite{kloe_pro}. The KLOE reach at full statistics for the $\pi\pi\gamma$ channel is shown in Fig.~\ref{Raw_TLimit_Output} at $N_{\rm U}/\sqrt{N_{\rm QED}}=2$ and a 2 MeV binning factor (green).
%\begin{figure}[htp!]
%\begin{center}
%\includegraphics[width=5.8cm]{fig7_acta_new.eps}
%\caption{Limits on $\varepsilon^2$ in the 0-1000 MeV range (see Fig.~\ref{Raw_TLimit_Output}), red line refers to kloe sensitivity at 2.5fb$^{-1}$ and $35^{\circ}<\theta_{\mu}<145^{\circ}$}
%\label{mmg_sensitivity}
%\end{center}
%\end{figure}

\section{Conclusions}
\label{conclusions}

We searched  for a light, dark vector boson through a study of the $\mu^+\mu^-\gamma$ ISR process by analysing a sample corresponding to a total integrated luminosity of 239~pb$^{-1}$. We found no evidence for such a $U$ boson. We set an upper limit at 90\% CL on the kinetic mixing parameter $\varepsilon^2$ between  \pt1.6;-5; and \pt8.6;-7; in the 520--980 MeV mass range. 
%The limit significantly improves the current limit on $\varepsilon$ in this mass range.
A future analysis that exploits full KLOE statistics and  an extended muon acceptance for the $\mu\mu\gamma$ channel as well as the investigation of $\pi\pi\gamma$ channel are planned.

\section*{Acknowledgments}
We warmly thank our former KLOE colleagues for the access to the data collected during the KLOE data taking campaign.
We thank the \DAF\ team for their efforts in maintaining low background running conditions and their collaboration during all data taking. We wish to thank our technical staff:
G.F. Fortugno and F. Sborzacchi for their dedication in ensuring efficient operation of the KLOE computing facilities;
M. Anelli for his continuous attention to the gas system and detector safety;
A. Balla, M. Gatta, G. Corradi and G. Papalino for electronics maintenance;
M. Santoni, G. Paoluzzi and R. Rosellini for general detector support;
C. Piscitelli for his help during major maintenance periods.
This work was supported in part by the EU Integrated Infrastructure Initiative Hadron Physics Project under contract number RII3-CT- 2004-506078; by the European Commission under the 7th Framework Programme through the `Research Infrastructures' action of the `Capacities' Programme, Call: FP7-INFRASTRUCTURES-2008-1, Grant Agreement No. 227431; by the Polish National Science Centre through the Grants No. 0469/B/H03/2009/37, 0309/B/H03/2011/40, DEC-2011/03/N/ST2/02641, \\
2011/01/D/ST2/00748, 2011/03/N/ST2/02652, 2013/08/M/ST2/00323 and by the Foundation for Polish Science through the MPD programme and the project HOMING PLUS BIS/2011-4/3.

%\subsection{Subsection}

%uncomment the following lines to place a figure
%\begin{figure}[htb]
%\centerline{%
%\includegraphics[width=12.5cm]{Fig1}}
%\caption{Plot of ...}
%\label{Fig:F2H}
%\end{figure}

\end{document}